# Decoding How Proteins Fold

Jorge A. Vila
IMASL-CONICET, Universidad Nacional de San Luis, Ejército de Los Andes 950, 5700 San Luis, Argentina.

One of the most puzzling and unsolved challenges in molecular biology is understanding *how* proteins fold. Despite having advanced predictive tools that can accurately estimate the native structures of proteins, we still lack a comprehensive model that explains how amino acid sequences dictate folding pathways and trajectories. This manuscript takes a fresh approach to this problem by resorting to the 'principle of least action.' This approach enables us to explore an intriguing question: *how* does a protein achieve its native state at a constant folding rate and within a time frame that is biologically plausible? A response to this inquiry will help us understand *why* proteins must fold along specific pathways and identify the boundary conditions that restrict their availability. It will also clarify *why* different folding pathways could be characterized by a common effective folding trajectory. Finally, it will provide a clear explanation for Levinthal's paradox. Our results are expected to pave the way for a more profound understanding of *how* proteins fold, shedding light on how the amino acid sequence and its surrounding environment encode the protein's folding pathways and, consequently, the protein's three-dimensional structure.

**Introduction**

The folding pathways problem originates from Levinthal (1968), who posed a paradox bearing his name: *how* can a protein attain its natural state within a biologically feasible timeframe while randomly exploring the whole conformational space? A proper answer to this query is of foremost importance since protein folds from milliseconds to seconds rather than ($\sim 10^{52}$) years—as foreseen by an exhaustive enumeration of all possible conformations for a 100-residue protein (Karplus, 2011). As the reader must be aware, several possible solutions to this apparent contradiction—also known as Levinthal's paradox—exist in the literature (Zwanzig *et al*., 1992; Karplus, 1997; Finkelstein & Badretdinov, 1977; Dill & Chan, 1997; Bogatyreva & Finkelstein, 2001; Rooman *et al*., 2002; Garbuzynskiy *et al*., 2012; Ben-Naim, 2012; Finkelstein & Garbuzynskiy, 2013; Martinez, 2014; Ivankov & Finkelstein, 2020; Vila, 2023a). Although we will not revisit each possible solution here, it is worth noting that solving this paradox is relevant because it could enable us to determine clear answers to the following key questions: *Why* and *how* can proteins reach their native state in a biologically reasonable time? Regarding the '*why*,' we demonstrated—from a thermodynamic viewpoint—that the range of 'slowest' folding times for two-state monomeric proteins ($\tau$) spans from ~milliseconds to ~seconds (Vila, 2023a), which closely aligns with the observed data (Garbuzynskiy *et al*., 2012; Ivankov & Finkelstein, 2020). These time scales for protein folding kinetics arise from the fulfillment of the thermodynamic hypothesis, also known as the Anfinsen dogma (Anfinsen, 1973), which allowed us to demonstrate—through a statistical-thermodynamics analysis (Vila, 2019)—the existence of an upper limit to the marginal stability of globular proteins ($\Delta G$) of approximately 7.4 kcal/mol (Vila, 2022), beyond which proteins either unfold or lose their functionality (Vila, 2021). This threshold applies to any fold class, sequence, or protein size, and it is significant because it arises from a quasi-equilibrium of forces that occurs at the lowest accessible protein's free-energy minimum (Martin & Vila, 2020). As to the '*how*'—which mainly focuses on clarifying the mechanisms by which proteins achieve their native states, including the pathways and trajectories of folding—it is a query that remains largely unresolved, despite the extensive literature available on the subject (Anfinsen & Scheraga, 1975; Creighton, 1985; Šali *et al*., 1994; Thirumalai, 1995; Baldwin, 1994; Baldwin, 1995; Wolynes *et al*., 1995; Lazaridis & Karplus, 1997; Dill & Chan, 1997; Pande *et al*., 1998; Dobson *et al*., 1998; Honig, 1999; Bakk *et al*., 2000; Englander *et al*., 2007; Karplus, 2007; Karplus, 2011; Englander & Mayne, 2014; Wolynes, 2015; Neupane *et al*., 2016; Lapidus, 2021;

Kikuchi, 2022; Zhao *et al*., 2023; Fakhoury *et al*., 2023; Zhao *et al.*, 2024; Chang & Perez, 2025). For a more profound understanding of the complexity of this problem, we should keep in mind that the protein folding process cannot be represented in a single dimension; rather, it results from interactions across multiple dimensions, creating complex multidimensional pathways and trajectories. This feature highlights the importance of understanding the interplay of complex forces at work, such as van der Waals interactions, hydrogen bonds, solvent effects, and electrostatic forces, all of which influence obtaining an accurate solution to the protein folding problem in a manner that demands solving it as an 'analytical whole' rather than a sum of parts (Vila, 2023b).

It is essential to acknowledge two key points prior to moving forward. Firstly, the terms 'pathway' and 'trajectory' represent distinct concepts. The difference between pathways and trajectories lies in the fact that pathways represent discrete sequences of steps to go from point A (*terminus a quo*) to point B (*terminus ad quem*), whereas trajectories encompass the duration of the process— as for the analysis of factors controlling protein evolvability (Vila, 2025). Secondly, it is well-documented experimentally that proteins can reliably attain the folded state at the same rate in a proper environment (Harrington & Schellman, 1956; Gromiha, 2005; Huang *et al*., 2008), thereby suggesting that *all* folding pathways—if multiple exist—may be characterized by an 'effective trajectory.' To gain a better grasp of this conjecture, we will analyze it from the standpoint of the 'least action principle,' a fundamental physical principle and a powerful model for describing how systems behave under the constraints mentioned above (Feynman *et al*., 1963; Hanc & Taylor, 2004). This principle's significance encompasses all domains of physics, rendering it the nearest approximation to a theory of everything, as it pertains to classical mechanics (Landau & Lifshitz, 1982), relativity (Landau & Lifshitz, 1987), quantum mechanics (Feynman, 2005), thermodynamics (Kaila & Annila, 2008), and biophysics (Simmons & Weiner, 2013). This approach will enable us to further investigate the more likely trajectory that proteins would take to reach their folded or unfolded states within a biologically reasonable timescale—and hence solve the Levinthal paradox—as well as uncover fundamental physical principles that dictate how life adapts and evolves in response to environmental or sequential changes.

**Outlining the system being studied**

The existence of metamorphic proteins (Murzin, 2008) indicates that the role of environmental factors cannot be overstated (Vila, 2020), as even small changes in conditions can lead to dramatically different folding outcomes (Dishman & Volkman, 2018). This viewpoint emphasizes that the dynamic nature of proteins means their behavior is not solely determined by intrinsic properties—encoded in their amino acid sequence (Anfinsen, 1973)—but also heavily influenced by external stimuli. This complexity highlights the necessity of studying protein behavior under varied conditions to understand the principles of folding and stability fully. A single-point change in the amino acid sequence—such as due to mutations—only exacerbates this problem by giving rise, among other things, to epistatic effects (Vila, 2024a).

In summary, based on the considerations mentioned above, we will now focus on analyzing the folding of a single two-state protein in a fixed environmental setting. The analysis of two-state proteins (Jackson, 1998) is chosen for a specific reason: the native state and unfolded states are separated by an energetic barrier that exceeds the energy of thermal fluctuations (Akmal & Muñoz, 2004; Kuwajima, 2020). In other words, the folded and unfolded states are separated by a collection of high-energy native-like structures known as the Transition State Ensemble (TSE), which represents (as shown in Figure 1) the energetic barrier (ΔG) for the process (Matouschek *et al*., 1989; Itzhaki *et al*., 1995; Englander, 2000; Ding *et al*., 2002; Vendruscolo *et al*., 2003; Akmal & Muñoz, 2004; Shakhnovich, 2006; Englander & Mayne, 2014; Lange *et al*., 2008; Kuwajima, 2020; Stiller *et al*., 2022; Li & Gong, 2022). In this simple folding model, there are no stable intermediate states necessary to complete the process. Therefore, the treatment of protein folding and unfolding can be viewed as interchangeable processes (Lazaridis & Karplus, 1997; Jackson, 1998; Ivankov & Finkelstein, 2020)—as long as the Gibbs free-energy barrier (ΔG) remains the same for both cases. The folding and unfolding data from 108 proteins, of which 70 display two-state kinetics (Glyakina & Galzitskaya, 2020), support these claims, as the logarithm of the folding and unfolding rates demonstrates a strong correlation ($R$ ~0.8). Henceforth, these terms shall be considered synonymous from this point forward.

**The least action principle and the effective-trajectory conjecture**

For a comprehensive and detailed exploration of the application of the principle of least action in classical mechanics, thermodynamics, relativity, and quantum mechanics, we recommend consulting the existing literature (Landau & Lifshitz, 1982; Landau & Lifshitz, 1987; Feynman, 2005; Kaila & Annila, 2008). We will skip over all those specific uses and instead provide, firstly, a layman's definition of the concept of least action and, later, a more concise definition of the principle from a physics-mathematics perspective.

The 'principle of least action' in physics states that a system going from a starting to an ending point—such as the folding process of proteins—will naturally follow a path that minimizes the 'action' (a functional of the trajectory) among all feasible pathways connecting these two points. The implication here is that nature operates with optimal efficiency, rather than randomly searching among all feasible pathways.

Let us now focus on a brief description of the principle of least action within the framework of Lagrangian mechanics—an approach that enables us to frame the system using a function encompassing both kinetic and potential energy contributions (Feynman *et al*., 1963; Hanc & Taylor, 2004). Let us begin by characterizing the folding of a protein as a process that occurs in a high-dimensional dihedral-angle space, which could be represented by a smooth, continuous function: $f(\{\varphi, \psi, \omega, \chi\}, t)$—where $\{\varphi, \psi, \omega, \chi\}$ denotes a set of internal dihedral angles of the protein at time *t*. This function determines how the protein conformation evolves along a given folding pathway under a given *all*-atom force field—which, in simple terms, refers to *all* the lines of force around and between atoms—such as the empirical conformational energy program for peptides (Arnautova *et al*., 2006). Additionally, we can define the time derivative of this function [$\dot{f}(\{\varphi, \psi, \chi\}, t)$], which identifies the folding velocity along a specified path that must be traversed within a well-defined range of folding rates ($\tau$). Specifically, it must satisfy the following boundary constraint: ~milliseconds < $\tau$ < ~seconds (Vila, 2023a)—as per an analysis grounded in transition state theory (Ivankov & Finkelstein, 2020)—with the folding rate calculated as $\tau = \tau_0 \exp(\beta \Delta G)$. Here, $\Delta G$ represents the protein's marginal stability (Vila, 2022), namely, the Gibbs free energy difference between the native state and the highest-energy native-like structures—of the transition state ensemble (TSE)—coexistent with it (see Figure 1); $\tau_0$ is a pre-exponential factor that indicates the folding speed limit of two-state proteins—also known as the barrier-less limit (Zana, 1975; McCammon, 1996; Mayor *et al*., 2000; Muñoz & Cerminara, 2016; Eaton, 2021)—and $\beta = 1/RT$, with *R* representing the gas constant and *T* being the absolute temperature. After providing some details of the folding process, the action (*S*), a functional of the trajectory, can be defined as the integral of the Lagrangian ($\mathcal{L}$) along any folding pathway, namely as:

$$S = \int_0^{\tau} \mathcal{L}\,[f(\varphi,\psi,\omega,\chi,t), \dot{f}(\varphi,\psi,\omega,\chi,t)]\;dt \qquad (1)$$

This path integral encompasses the entire duration of the folding process, under the constraint that the upper-bound limit of the integration ($\tau$) must be less than a few seconds. This cap on the slowest folding time—for any two-state protein, regardless of its fold class, length, or sequence—was proven to arise from the validity of the thermodynamic hypothesis, or Anfinsen's dogma (Vila, 2023a). It is worth noting that this analysis also enables us to explain *why* protein folding consistently occurs within a biologically reasonable time frame. In summary, the validity of the Anfinsen dogma specifically sets an integration limit that confines feasible folding pathways.

The task of solving equation (1) presents an enormous mathematical challenge that exceeds the current analysis's limits, as it requires solving—among other things—the Euler-Lagrange equations [ $\frac{d}{dt}\left(\frac{d\mathcal{L}}{d\dot{f}}\right) - \left(\frac{d\mathcal{L}}{df}\right) = 0$] that represent the solution to virtual variations on the functional *S* (Coopersmith, 2017); instead, we will focus on highlighting its main features, aiming to clarify the potential relationship between the effective-trajectory conjecture and the least action principle. To achieve this, it is important to note that the Hamilton principle (of stationary action) states that the variations ($\delta$) that make the integral zero ($\delta S = 0$), by infinitesimal virtual variation on *f*, correspond to a stationary point of *S*—where the functional must be at a minimum or a saddle point, but not a maximum. All things considered, a pathway is considered a stationary point of the action (*S*) if—and only if—the Euler-Lagrange equations are satisfied (Coopersmith, 2017). Hence, protein folds by following a pathway that renders the action (*S*) stationary. However, how can we guarantee that the action has a unique path that makes it stationary? The latter represents an important question, as the number of folding pathways—specifically, whether there is one, a few, or a countless number—has been, and continues to be, a topic of debate (Baldwin, 1994; Baldwin, 1995; Lazaridis & Karplus, 1997; Bakk *et al*., 2000; Rooman *et al*., 2002; Englander & Mayne, 2014; Eaton & Wolynes, 2017; Englander & Mayne, 2017). Then, what if numerous pathways exist, each represented by trajectories that share the same folding rate ($\tau$) as their sole common feature? This viewpoint associates the uniqueness of the solution to Eq. (1) with the time necessary for the protein to navigate the most efficient pathways—those that render the action (*S*) stationary—rather than suggesting the existence of a specific one (refer to Figure 2). This leads to the conjecture of an ‘effective trajectory,’ which encompasses *all* folding pathways characterized by a common single folding rate ($\tau_{single}$), depending on the specific conditions described below. Firstly, the folding process must take place in a proper and controlled environment, which includes factors such as pH, temperature, and ionic strength, among others. Secondly, the single-folding rate must satisfy the following constraint: $\tau_{single} < \sim$seconds, as previously rationalized.

The proposed effective-trajectory conjecture is based on a crucial observation: proteins unfold at a single-folding rate determined by ΔG (Vila, 2023a), which defines the protein's marginal stability while also serving as a threshold beyond which a protein may unfold or lose its functionality (refer to Figure 1). This scenario offers a simple and credible explanation for how multiple pathways may converge on the protein native state at the same rate (see Figure 2)—even in the presence of a degenerate Gibbs free energy minimum (see Figure 3), a possibility that cannot be ruled out (Vila *et al*., 2003; Martin *et al*., 2019). Regardless of the latter, it is important to recognize that different proteins—with varying sequences, numbers of amino acids, or both—can display identical folding rates. Proteins with Protein Data Bank identifiers 1PBA, 1AYE, and 2VIK illustrate this phenomenon (Huang *et al*., 2008). At first glance, it may appear odd that such proteins can exhibit identical folding rates while possessing distinct folding pathways. However, upon recognizing two pivotal facts, this phenomenon becomes readily comprehensible. Firstly, the folding rates are limited—as previously mentioned—by the protein's marginal stability—regardless of fold class, sequence, or protein size (Vila, 2023a). This proposal is consistent with evidence suggesting that folding rates are insensitive to a protein's sequence details (Plaxco *et al*., 2000). Secondly, paths and trajectories are not—as already noted—synonymous: a path denotes a sequence of events or mechanisms that links the starting and ending points of a process, such as protein folding. On the other hand, a trajectory accounts for the time span—as provided by the folding rate—required for the protein to navigate such a path. The latter clarification is of paramount importance, as the thermodynamic hypothesis, or Anfinsen dogma (1973), proposes that the amino acid sequence determines, in a proper environment, a protein's unique three-dimensional structure—a statement that remains valid even for metamorphic proteins (Vila, 2020)—regardless of how long it takes for the protein to fold or unfold. In other words, the amino acid sequence and its surrounding environment encode its folding pathways and consequently the protein's structure.

All in all, for more than half a century, chemists, biologists, mathematicians, and physicists have been unable to solve the protein folding problem analytically (Vila, 2023b). Consequently, as an alternative, we have employed a heuristic argument to both demonstrate the physical plausibility of the effective-trajectory conjecture and to illustrate the factors controlling the feasible folding pathways in a more intuitive manner, *i.e.*, without relying on exhaustive mathematical rigor. The upcoming section will examine the validity range of the proposed solution, its relation to the Levinthal paradox, and its implications on critical issues in structural and evolutionary biology.

**Avenues for future research.**

Knowledge of the single folding rate ($\tau_{single}$) that characterizes the effective-trajectory solution of Eq. (1)—for a given environment—is crucial in practical applications. For example, it allows researchers to more precisely forecast changes in protein stability upon varying initial conditions (Vila, 2024b), such as those arising from single-point mutations (Vila, 2022). This analysis does not undermine the significance of understanding the pathway a protein could take, as this knowledge offers vital information regarding the key factors affecting protein evolvability (Vila, 2025). Unfortunately, as already mentioned, resolving this issue requires an analytical solution to Eq. (1), which—as previously indicated (Vila, 2023b)—presents a major unsolved challenge. This nuanced comprehension highlights the complexity of protein folding dynamics, suggesting that while multiple pathways might exist, all of them ultimately converge at a common destination: the native state, which must be attained at a given single folding rate ($\tau_{single}$)—as determined by $\Delta G_{single}$ (as shown in Figure 2). This convergence strengthens the idea that the folding process can be simplified—for practical applications—into an 'effective trajectory.' The latter aligns with the thorough analysis of the unfolding pathways of *chymotrypsin inhibitor 2* conducted by Lazaridis and Karplus (1997). Their simulations indicate that a "… *statistically preferred unfolding pathway*…" is feasible, providing evidence that supports the proposed single effective-trajectory conjecture.

The whole analysis enables us, on the one hand, to bridge the effective-trajectory conjecture—originating from the validity of the least action principle—with the phenomenon of multiple pathways. This is a critical point, as it does not contradict the 'new view' of protein folding (Baldwin, 1994; Lazaridis & Karplus, 1997), which posits that there are multiple paths to the native state. On the other hand, it solves the Levinthal paradox (regarding the folding search timescale), as only those pathways exhibiting a common single folding rate ($\tau_{single}$)—as provided by the least action principle—are allowed during the conformational search. This hypothesis dismantles the idea that the conformational search for the native state could be random. From this perspective, the paradox vanishes because each feasible protein folding pathway represents a solution.

We have thus far examined *how* proteins attain their native states regarding pathways and trajectories, but we have not addressed the mechanisms underlying that conformational search—such as the 'funnel model' (Dill & Chan, 1997) or the 'foldon model' (Bai *et al*., 1995)—as this falls outside the primary focus of our research. It is critical, however, to recognize these mechanisms' significance for comprehending a wide range of biological processes and illnesses. Thus, studying how these mechanisms affect protein pathways and trajectories may improve our understanding of protein folding and misfolding, which should be considered in future research.

Overall, future research in structural and evolutionary biology should prioritize solving Eq. (1). By solving this equation, researchers will gain crucial information that deepens our understanding of the main factors influencing protein folding and misfolding, as well as the feasible folding pathways. This knowledge will enable us to precisely and accurately determine *how* the amino acid sequence encodes its folding and, hence, which mechanism better represents the protein folding process.

## Conclusions

In this work, we provide a reasonable answer to a critical question in evolutionary biology: *how* a protein always attains—through a single or multiple folding pathways—its native state within a biologically acceptable timeframe. We have used the 'least action principle' as a physical framework to rationalize this process, *i.e*., to disclose *why* proteins must fold by following specific pathways—those that make the 'action' stationary. All these specific pathways have in common a unique folding rate determined mainly by the protein's marginal stability ($\Delta G$)—which ultimately is set by the equilibrium of native and native-like conformations populating the transition state ensemble. This indicates that if multiple pathways are present, they must all be characterized by the same $\Delta G$ to ensure proteins fold at the most efficient rate, regardless of which path nature chooses to reach the native state. This proposal implies that proteins will achieve the native state within a biologically acceptable timeframe. The latter is assured by the validity of the thermodynamic hypothesis—also known as the Anfinsen dogma—which determines a constraint for the slowest single-folding rate ($\tau_{single} < \sim$seconds). In general, the principle of least action should be recognized as a fundamental physical law that allows us to understand *how* proteins fold in accordance with rigorous physical-mathematical rules and specific boundary conditions—such as limiting the folding pathways to those that render the 'action' stationary and whose slowest folding rate is capped at a few seconds.

The analysis also revealed that Levinthal's paradox should never have been considered as a challenge to *how* proteins fold. It would have been enough to assume—even without knowing exactly what all the forces are and how they collectively act during the folding process—that nature will always follow the most efficient folding pathway, rather than search for it randomly, as determined by the universal 'principle of least action'. Even if countless pathways satisfied this principle, there would be no conformational-search timescale paradox, as each would reach its native state within a biologically reasonable timeframe. The real challenge is understanding the mechanisms that lead proteins to fold into their native state, and this question remains unanswered after more than 50 years of research in the structural biology field. The reason for these difficulties is that the central question of the protein folding problem remains unresolved: specifically, *how* a sequence of amino acids, in a proper environment, encodes its folding pathways. However, advancements in the state of the art of computational modeling, simulations, and experimental techniques are gradually shedding light on what these pathways could be, providing optimism for potential breakthroughs in our understanding of such essential biological processes.

**Acknowledgement**

The author acknowledges support from the Institute of Applied Mathematics San Luis (IMASL), the National University of San Luis (UNSL), and the National Research Council of Argentina (CONICET). I want to extend my gratitude to Dr. Pablo Garay for his assistance in preparing Figure 3.

**Founding**

This research did not receive any specific grant from funding agencies in the public, commercial, or not-for-profit sectors.

**References**

Akmal A, Muñoz V. The nature of the free energy barriers to two-state folding. *Proteins*. 2004, 57, 142-52.

Anfinsen CB, Scheraga HA. Experimental and theoretical aspects of protein folding. Adv Protein Chem 1975, 29, 205-300.

Anfinsen CB. Principles that govern the folding of protein chains. Science 1973, 181, 223-230.

Arnautova YA, Jagielska A, Scheraga HA. A new force field (ECEPP-05) for peptides, proteins, and organic molecules. J Phys Chem B 2006, 110(10), 5025-5044.

Bai Y, Sosnick TR, Mayne L, Englander SW. Protein folding intermediates: native-state hydrogen exchange. Science 1995, 269(5221), 192-197.

Bakk A, Høye JS, Hansen A, Sneppen K, Jensen MH. Pathways in two-state protein folding. Biophys J 2000, 79(5), 2722-2727.

Baldwin RL. Protein folding. Matching speed and stability. Nature 1994, 369(6477):183-4.

Baldwin RL. The nature of protein folding pathways: the classical versus the new view. J Biomol NMR 1995, 5(2), 103-109.

Ben-Naim A. Levinthal's Paradox Revisited and Dismissed. Open Journal of Biophysics 2012, 2, 23-32.

Bogatyreva NS, Finkelstein AV. Cunning simplicity of protein folding landscapes. Protein Eng 2001, 14(8), 521-523.

Chang L, Perez A. Rapid estimation of protein folding pathways from sequence alone using AlphaFold2. Nat Commun(2025. https://doi.org/10.1038/s41467-025-66870-x

Coopersmith, J. The Lazy Universe: An Introduction to the Principle of Least Action. Oxford University Press (2017), https://doi.org/10.1093/oso/9780198743040.003.0001.

Creighton TE. The problem of how and why proteins adopt folded conformations. The Journal of Physical Chemistry 1985, 89(12), 2452-2459.

Dill KA, Chan HS. From Levinthal to pathways to funnels. Nat Struct Biol 1997, 4(1), 10-19.

Ding F, Dokholyan NV, Buldyrev SV, Stanley HE, Shakhnovich EI. Direct molecular dynamics observation of protein folding transition state ensemble. Biophys J 2002, 83(6), 3525-3532.

Dishman AF, Volkman BF. Unfolding the Mysteries of Protein Metamorphosis. ACS Chemical Biology 2018, 13(6), 1438-1446.

Dobson CM, Šali A, Karplus M. Protein Folding: A Perspective from Theory and Experiment. Angew Chem Int Ed Engl 1998, 37(7), 868-893.

Eaton WA, Wolynes PG. Theory, simulations, and experiments show that proteins fold by multiple pathways. Proc Natl Acad Sci USA 2017, 114(46), E9759-E9760.

Eaton WA. Modern Kinetics and Mechanism of Protein Folding: A Retrospective. J Phys Chem B 2021, 125, 3452-3467.

Englander SW, Mayne L, Krishna MM. Protein folding and misfolding: mechanism and principles. Q Rev Biophys 2007, 40(4), 287-326.

Englander SW, Mayne L. Reply to Eaton and Wolynes: How do proteins fold? Proc Natl Acad Sci USA. 2017, 114(46), E9761-E9762.

Englander SW, Mayne L. The nature of protein folding pathways. Proc Natl Acad Sci USA 2014, 111(45), 15873-15880.

Englander SW. Protein folding intermediates and pathways studied by hydrogen exchange. Annu Rev Biophys Biomol Struct 2000, 29, 213-38.

Fakhoury Z, Sosso GC, Habershon S. Generating Protein Folding Trajectories Using Contact-Map-Driven Directed Walks. J Chem Inf Model 2023, 63(7), 2181-2195.

Feynman RP, Leighton RB, and Sands M, *The Feynman Lectures on Physics* (Addison-Wesley, Reading, MA, 1963), Vol. II, Chap. 19.

Feynman RP. The Principle of Least Action in Quantum Mechanics. Feynman's Thesis—A New Approach to Quantum Theory; World Scientific, 2005; pp 1-69.

Finkelstein AV, Badretdinov AYa. Rate of protein folding near the point of thermodynamic equilibrium between the coil and the most stable chain fold. Fold Des 1997, 2(2), 115-121.

Finkelstein AV, Garbuzynskiy SO. Levinthal's question answered … again? J Biomol Struct Dyn. 2013, 31(9), 1013-1015.

Garbuzynskiy SO, Ivankov DN, Bogatyreva NS, Finkelstein AV. Golden triangle for folding rates of globular proteins. Proc Natl Acad Sci USA 2013, 110(1), 147-50.

Glyakina AV, Galzitskaya OV. How Quickly Do Proteins Fold and Unfold, and What Structural Parameters Correlate with These Values? Biomolecules 2020, 10, 197.

Gromiha MM. A statistical model for predicting protein folding rates from amino acid sequence with structural class information. J Chem Inf Model 2005, 45(2), 494-501.

Hanc J, Taylor EF. From conservation of energy to the principle of least action: a story line. Am J Phys 2004, 72, 514–521.

Harrington WF, Schellman JA. Evidence for the instability of hydrogen-bonded peptide structures in water, based on studies of ribonuclease and oxidized ribonuclease. C R Trav Lab Carlsberg Chim 1956, 30(3), 21-43.

Honig B. Protein folding: from the Levinthal paradox to structure prediction. J Mol Biol 1999, 293(2), 283-293.

Huang LT, Gromiha MM. Analysis and prediction of protein folding rates using quadratic response surface models. J Comput Chem 2008, 29(10), 1675-1683.

Itzhaki LS, Otzen DE, Fersht AR. The structure of the transition state for folding of chymotrypsin inhibitor 2 analyzed by protein engineering methods: evidence for a nucleation-condensation mechanism for protein folding. J Mol Biol 1995, 254, 260-288.

Ivankov DN, Finkelstein AV. Solution of Levinthal's Paradox and a Physical Theory of Protein Folding Times. Biomolecules 2020, 10(2), 250.

Jackson SE. How do small single-domain proteins fold? Fold Des 1998, 3(4), R81-91.

Kaila VRI, Annila A. Natural selection for least action. Proc R Soc A 2008, 464, 3055-3070.

Karplus M. Behind the folding funnel diagram. Nat Chem Biol 2011, 7(7), 401-404.

Karplus M. The Levinthal paradox: yesterday and today. Fold Des 1997, 2(4), S69-75.

Kikuchi T. Decoding an Amino Acid Sequence to Extract Information on Protein Folding. Molecules 2022, 27(9), 3020.

Kuwajima K. The Molten Globule, and Two-State vs. Non-Two-State Folding of Globular Proteins. Biomolecules 2020, 10, 407.

Landau LD, Lifshitz EM. Course of Theoretical Physics Volume 1: Mechanics, 3rd ed.; Butterworth-Heinemann: Oxford, 1982.

Landau LD, Lifshitz EM. Course of Theoretical Physics Volume 2: The Classical Theory of Fields, 4th ed.; Butterworth-Heinemann: Oxford, 1987.

Lange OF, Lakomek N-A, Farès C, Schröder GF, Walter KFA, Becker S, Meiler J, Grubmüller H, Griesinger C, de Groot BL. Recognition Dynamics Up to Microseconds Revealed from an RDC-Derived Ubiquitin Ensemble in Solution. Science 2008, 13, 1471-1475.

Lapidus LJ. The road less traveled in protein folding: evidence for multiple pathways. Curr Opin Struct Biol 2021, 66, 83-88.

Lazaridis T, Karplus M. 'New view' of protein folding reconciled with the old through multiple unfolding simulations. Science 1997, 278(5345), 1928-1931.

Levinthal C. Are There Pathways for Protein Folding? Journal de Chimie Physique 1968, 65, 44-45.

Li Y, Gong H. Identifying a Feasible Transition Pathway between Two Conformational States for a Protein. J Chem Theory Comput 2022, 18, 4529-4543.

Martin AO, Vila JA. The Marginal Stability of Proteins: How the Jiggling and Wiggling of Atoms is Connected to Neutral Evolution. Journal of Molecular Evolution 2020, 88, 424-426.

Martin OA, Vorobjev Y, Scheraga HA, Vila JA. Outline of an experimental design aimed to detect a protein A mirror image in solution. PeerJ Phys Chem 2019, 1, e2.

Martinez L. Introducing the Levinthal's Protein Folding Paradox and Its Solution. J Chem Educ 2014, 91, 11, 1918–1923.

Matouschek A, Kellis JT Jr, Serrano L, Fersht AR. Mapping the transition state and pathway of protein folding by protein engineering. Nature 1989, 340, 122-126.

Mayor U, Johnson CM, Daggett V, Fersht AR. Protein folding and unfolding in microseconds to nanoseconds by experiment and simulation. Proc Natl Acad Sci USA 2000, 97, 13518-22.

McCammon JA, Gelin BR, Karplus M. Dynamics of folded proteins. Nature 1977, 267, 585-590.

McCammon JA. A speed limit for protein folding. Proc Natl Acad Sci USA 1996, 93,11426-11427.

Muñoz V, Cerminara M. When fast is better: protein folding fundamentals and mechanisms from ultrafast approaches. Biochem J 2016, 473, 2545-2559.

Murzin, A.G. Biochemistry. Metamorphic proteins. Science 2008, 320, 1725-1726.

Neupane, K., Manuel, A. & Woodside, M. Protein folding trajectories can be described quantitatively by one-dimensional diffusion over measured energy landscapes. Nature Phys 2016, 12, 700-703.

Pande VS, Grosberg AY, Tanaka T, Rokhsar DS. Pathways for protein folding: is a new view needed? Curr Opin Struct Biol 1998, 8(1), 68-79.

Plaxco KW, Simons KT, Ruczinski I, Baker D. Topology, stability, sequence, and length: defining the determinants of two-state protein folding kinetics. Biochemistry 2000 (37), 11177-11183.

Rooman M, Dehouck Y, Kwasigroch JM, Biot C, Gilis D. What is paradoxical about Levinthal paradox? J Biomol Struct Dyn 2002, 20(3), 327-329.

Šali A, Shakhnovich E, Karplus M. How does a protein fold? Nature 1994, 369(6477), 248-251.

Shakhnovich E. Protein folding thermodynamics and dynamics: where physics, chemistry, and biology meet. Chem Rev 2006, 106, 1559-1588.

Simmons W, Weiner JL. The principle of stationary action in biophysics: stability in protein folding. J Biophys 2013, 2013:697529.

Stiller JB, Otten R, Häussinger D, Rieder PS, Theobald DL, Kern D. Structure determination of high-energy states in a dynamic protein ensemble. Nature. 2022, 603(7901), 528-535.

Thirumalai D. From Minimal Models to Real Proteins: Time Scales for Protein Folding Kinetics. Journal de Physique I 1995, 5 (11), 1457-1467.

Tsuboyama K, Dauparas J, Chen J, Laine E, Mohseni Behbahani Y, Weinstein JJ, Mangan NM, Ovchinnikov S, Rocklin GJ. Mega-scale experimental analysis of protein folding stability in biology and design. Nature 2023, 620(7973), 434-444.

Vendruscolo M, Paci E, Dobson CM, Karplus M. Rare Fluctuations of Native Proteins Sampled by Equilibrium Hydrogen Exchange. J Am Chem Soc 2003, 125(51), 15686-15687.

Vila JA, Ripoll DR, Scheraga HA. Atomically detailed folding simulation of the B domain of staphylococcal protein A from random structures. Proc Natl Acad Sci USA 2003, 100(25), 14812-14816.

Vila JA. Analysis of proteins in the light of mutations. Eur Biophys J 2024b, 53(5-6), 255-265.

Vila JA. Factors controlling protein evolvability—at the molecular scale. Eur Biophys J 2025 (*in press*).

Vila JA. Forecasting the upper bound free energy difference between protein native-like structures. Physica A 2019, 533, 122053.

Vila JA. Metamorphic Proteins in Light of Anfinsen's Dogma. J Phys Chem Lett 2020, 11(13), 4998-4999.

Vila JA. Protein Evolution upon Point Mutations. ACS Omega 2022, 7, 14371-14376.

Vila JA. Protein folding rate evolution upon mutations. Biophys Rev 2023a, 15, 661-669.

Vila JA. Rethinking the protein folding problem from a new perspective. Eur Biophys J 2023b, 52(3), 189-193.

Vila JA. The origin of mutational epistasis. Eur Biophys J 2024a, 53, 473-480.

Vila JA. Thoughts on the Protein's Native State. J Phys Chem Lett 2021, 12, 5963-5966.

Wolynes PG, Onuchic JN, Thirumalai D. Navigating the folding routes. Science 1995, 267(5204), 1619-1620.

Wolynes PG. Evolution, energy landscapes, and the paradoxes of protein folding. Biochimie 2015, 119, 218-230.

Zana R. On the rate determining step for helix propagation in the helix-coil transition of polypeptides in solution. Biopolymers 1975, 14, 2425-2428.

Zhao K, Xia Y, Zhang F, Zhou X, Li SZ, Zhang G. Protein structure and folding pathway prediction based on remote homologs recognition using PAthreader. Commun Biol 2023, 6(1), 243.

Zhao K, Zhao P, Wang S, Xia Y, Zhang G. FoldPAthreader: predicting protein folding pathway using a novel folding force field model derived from the known protein universe. Genome Biol 2024, 25(1), 152.

Zwanzig R, Szabo A, Bagchi B. Levinthal's paradox. Proc Natl Acad Sci USA 1992, 89, 20-22.

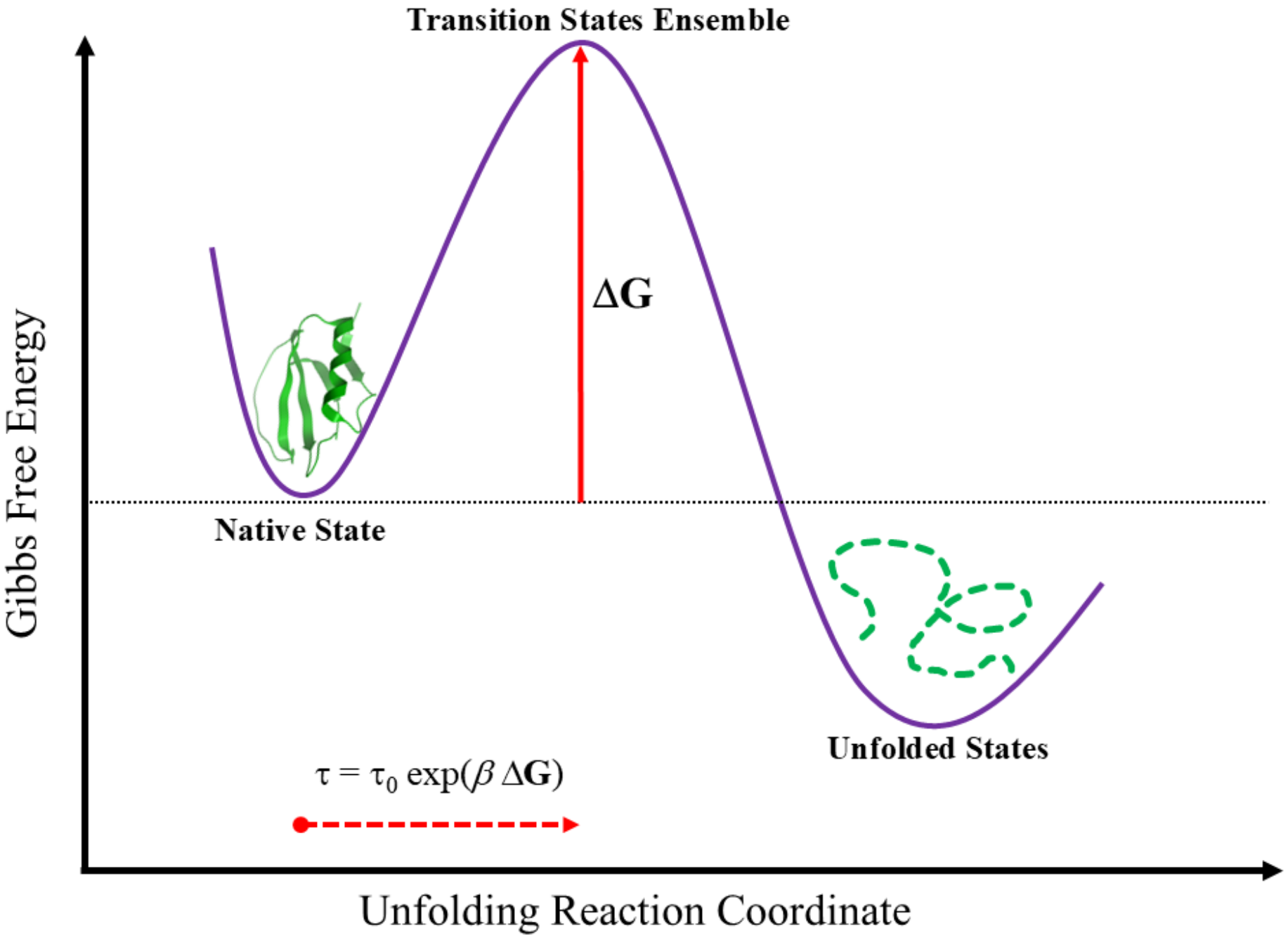

**Figure 1.** The Gibbs free-energy profile for the unfolding of a two-state protein is presented in an oversimplified format. The native state is illustrated by a green-ribbon diagram of a protein, while the highest point of the free-energy profile represents the Transition States Ensemble—a high-energy set of structures that coexist in fast dynamic equilibrium with the native state (McCammon *et al*., 1977; Vendruscolo *et al*., 2003; Lange *et al*., 2008). The Gibbs free-energy difference between these two states is represented by ΔG, indicating the protein's marginal stability (Martin & Vila, 2020)—a point beyond which the protein unfolds or becomes nonfunctional. The unfolding rate (τ) for a two-state protein—derived from first-principles calculations (Vila, 2023a)—is inset in the figure. For additional details, please consult the main text.

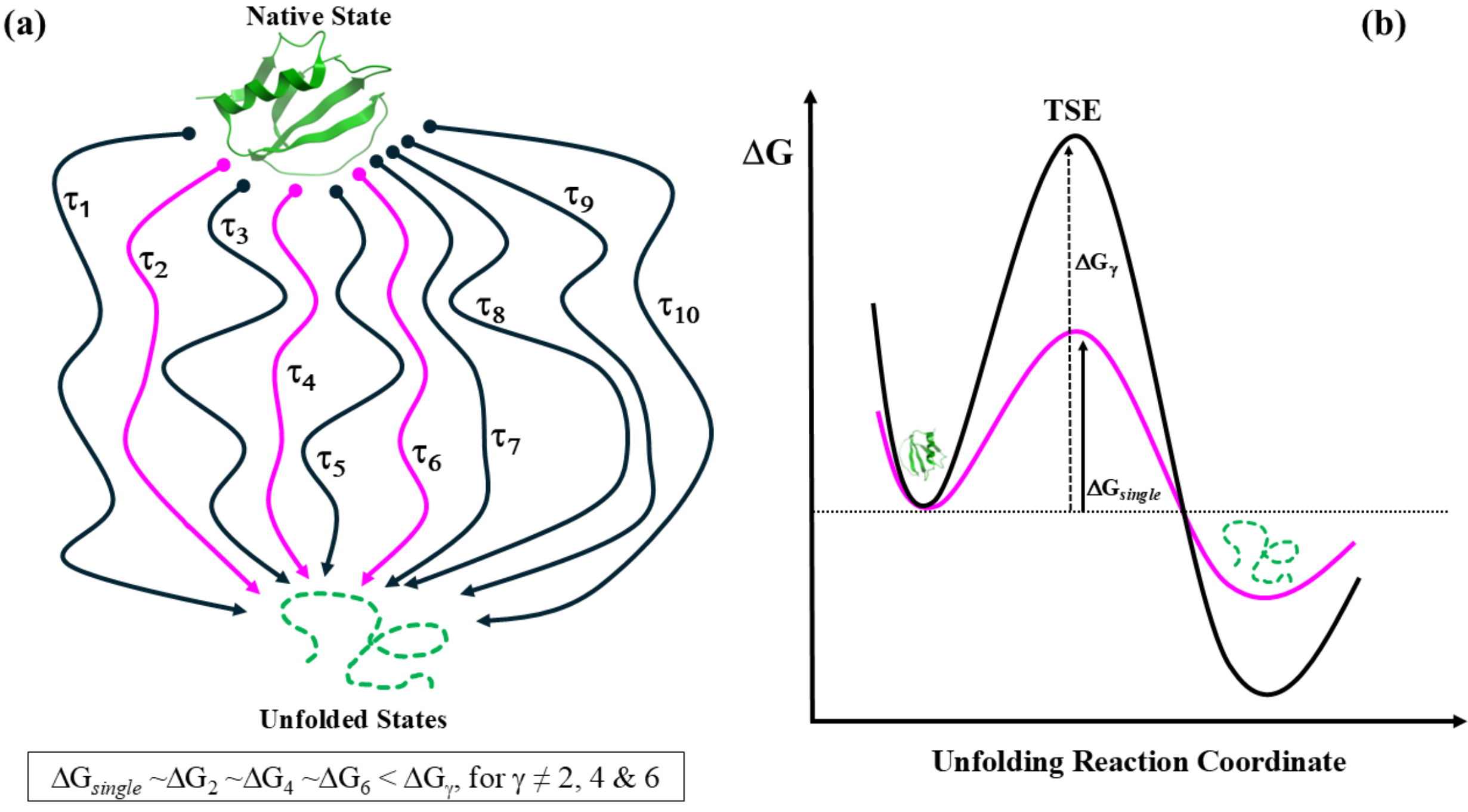

**Figure 2.** Panel (**a**) illustrates some pathways, among the many possible unfolding routes, in a cartoon format. Thick (black and magenta) lines depict these pathways, which begin at the native state and conclude at the unfolded states. The speed at which each pathway is traversed—during the unfolding process— determines the unfolding rate ($\tau$) for the corresponding trajectory, which is given by $\tau_x = \tau_0 \exp(\beta\, \Delta G_x)$, with $x = 1$ to 10. The figure also shows—in panel (**a**)—an arbitrary distribution of the Gibbs free-energy differences ($\Delta G_x$) associated with each pathway. Three of the ten trajectories, namely those that align with a given Gibbs free energy ($\Delta G_{single}$) value—defined by the folding rate resulting from a pathway that complied with the principle of least action—are highlighted in magenta (see main text for further details). The chosen arbitrary distribution for ΔG's in panel (**a**) mirrors the group of higher-energy-native-like structures populating the transition states ensemble (TSE)—as shown in panel (**b**).

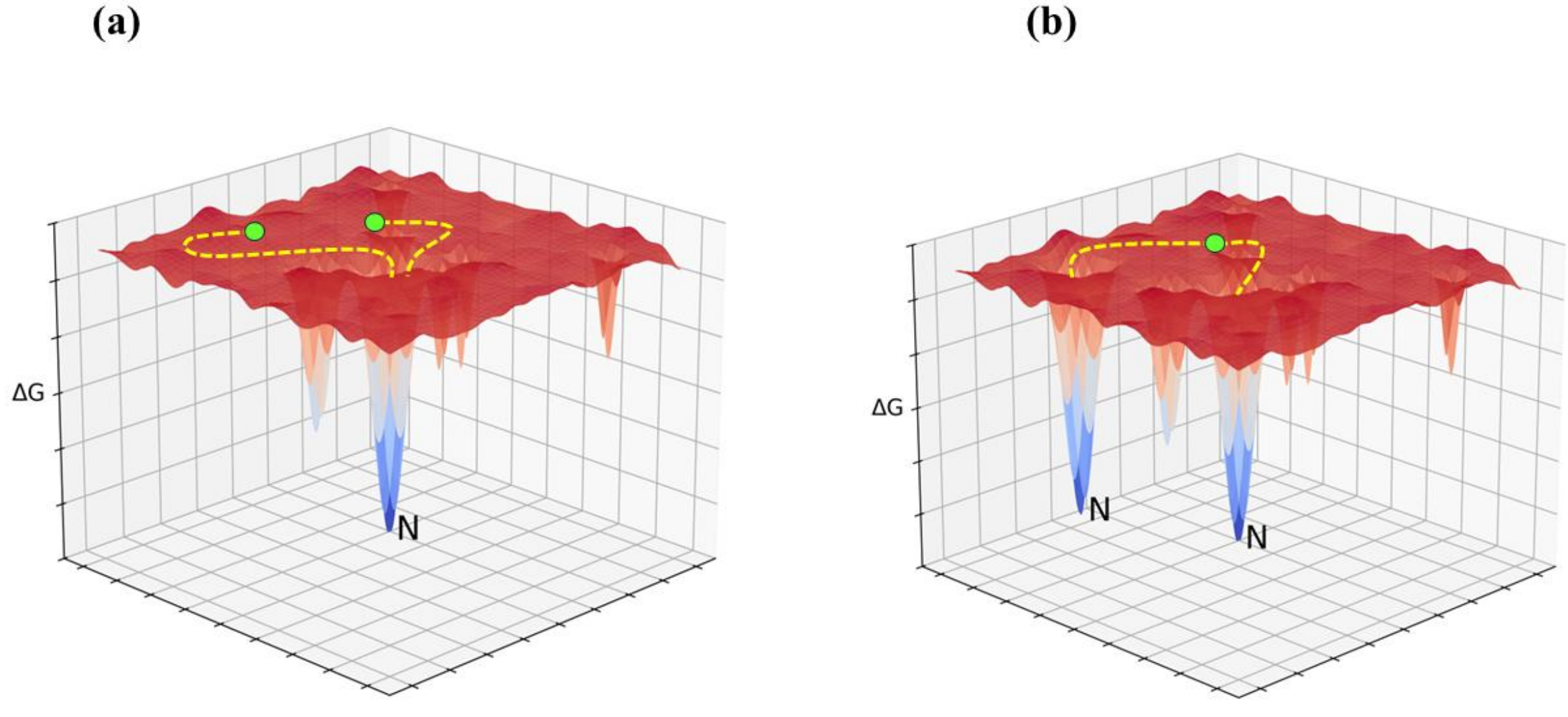

**Figure 3.** Panel (**a**) illustrates the Gibbs free-energy landscape for a two-state protein in a cartoon format. The lowest accessible free-energy minimum associated with the protein's native state is denoted as **N**. Panel (**b**) illustrates the scenario in which the lowest accessible free energy minimum is degenerate—a possibility that cannot be dismissed (Vila *et al*., 2003; Martin *et al*., 2019). In both panels, the green-filled dots highlight an arbitrariily selected starting point, and the dashed yellow lines delineate potential pathways ending at the native state (**N**).